\documentclass[12pt]{article}
\usepackage{graphicx}

\def\pbnr{}
\def\speaker{Kai-Thomas Brinkmann}
\def\onbehalfof{the PANDA collaboration}
\def\title{Status and Perspectives of $\bar{P}ANDA$}
\def\affiliation{II. Physikalisches Institut\\
Justus-Liebig-Universit\"{a}t Gie\ss{}en, Germany}
\def\support{Work supported by BMBF and $HICforFAIR$. $\bar{P}ANDA$ is supported by the national funding agencies of the participating groups.}
\newcommand\pubnumber{\pbnr}
\newcommand\pubdate{\today}
\def\Title#1{\begin{center} {\Large #1 } \end{center}}
\def\Author#1{\begin{center}{ \sc #1} \end{center}}

\newcommand{\OnBehalf}[1]{\sbox0{#1}\ifdim\wd0=0pt
        {}% if #1 is empty
	\else
	{\\on behalf of #1}% if #1 is not empty
	\fi}
\newcommand{\SupportedBy}[1]{\sbox0{#1}\ifdim\wd0=0pt
        {}% if #1 is empty
	\else
	{\footnote{#1}}% if #1 is not empty
	\fi}
\def\Address#1{\begin{center}{ \it #1} \end{center}}

\newcommand\pubblock{\includegraphics[width=5cm]{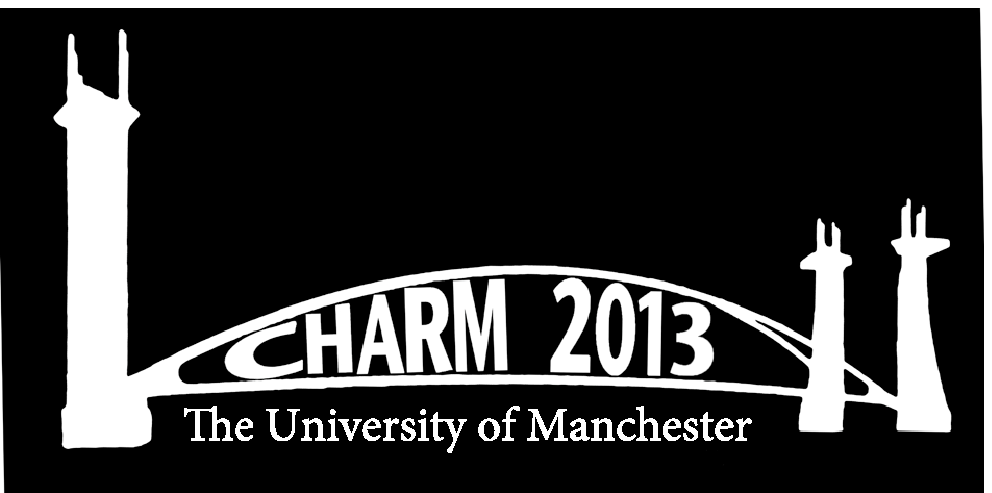}\hfill{\begin{tabular}{l} \pubnumber\\
         \pubdate  \end{tabular}}}
\newenvironment{Abstract}{\begin{quotation}  }{\end{quotation}}
\newenvironment{Presented}{\begin{quotation} \begin{center} 
             PRESENTED AT\end{center}\bigskip 
      \begin{center}\begin{large}}{\end{large}\end{center} \end{quotation}}
\def\Acknowledgements{\bigskip  \bigskip \begin{center} \begin{large}
             \bf ACKNOWLEDGEMENTS \end{large}\end{center}}
\def\venue{The 6$^{th}$ International Workshop on Charm Physics\\
(CHARM 2013)\\
Manchester, UK,  31 August -- 4 September, 2013}
\def\beq{\begin{equation}}
\def\eeq#1{\label{#1}\end{equation}}
\def\eeqn{\end{equation}}

\def\beqa{\begin{eqnarray}}
\def\eeqa#1{\label{#1}\end{eqnarray}}
\def\eeqan{\end{eqnarray}}

\let\bar=\overbar

\def\Dslash{\not{\hbox{\kern-4pt $D$}}}
\def\dslash{\not{\hbox{\kern-2pt $\del$}}}

\def\msb{{\bar{\ssstyle M \kern -1pt S}}}

\begin{document}
\begin{titlepage}
\pubblock

\vfill
\Title{\title}
\vfill
\Author{\speaker\SupportedBy{\support}\OnBehalf{\onbehalfof}}
\Address{\affiliation}
\vfill
\begin{Abstract}
%%%%%%%%%%%%%%%%%%%%%%%%%%%%%%%%%%%%%%%%%%%%%%%%%%%%%%%%%%%%%%%%%%%%%%%%%%%
% YOUR ABSTRACT GOES HERE
%%%%%%%%%%%%%%%%%%%%%%%%%%%%%%%%%%%%%%%%%%%%%%%%%%%%%%%%%%%%%%%%%%%%%%%%%%%
Physics with antiprotons in the charmonium mass region will play a major role at the future  $\bar{P}ANDA$ experiment at the $FAIR$ facility in Darmstadt. At $\bar{P}ANDA$, an antiproton beam with momenta up to 15 GeV/c circulating in the high-energy storage ring HESR will interact with a hydrogen target. High interaction rates and unprecedented momentum precision will allow experiments addressing, among a variety of other physics goals, hidden and open charm spectroscopy. The detector system of $\bar{P}ANDA$ is optimized to meet the challenges of high-resolution spectroscopy of charmonium states of any quantum number in formation and production with very good background suppression. At the same time, emphasis is placed on meeting the requirements of other parts of the physics program as, e.g. spectroscopy of hypernuclei, with a flexible setup of detector components. 
This contribution discusses the layout of the $\bar{P}ANDA$ detector. Apart from a description of the overall layout of the experiment, several novel detector developments will be described. The physics reach of the  $\bar{P}ANDA$ experiment in the charmonium sector is discussed in~\cite{SL13}.
\end{Abstract}
\vfill
\begin{Presented}
\venue
\end{Presented}
\vfill
\end{titlepage}
\def\thefootnote{\fnsymbol{footnote}}
\setcounter{footnote}{0}
%

%%%%%%%%%%%%%%%%%%%%%%%%%%%%%%%%%%%%%%%%%%%%%%%%%%%%%%%%%%%%%%%%%%%%%%%%%%%
%  WHAT FOLLOWS IS YOUR TEXT
%%%%%%%%%%%%%%%%%%%%%%%%%%%%%%%%%%%%%%%%%%%%%%%%%%%%%%%%%%%%%%%%%%%%%%%%%%%
\section{Introduction}

\begin{figure}[htb]
\centering
\includegraphics[height=3.0in]{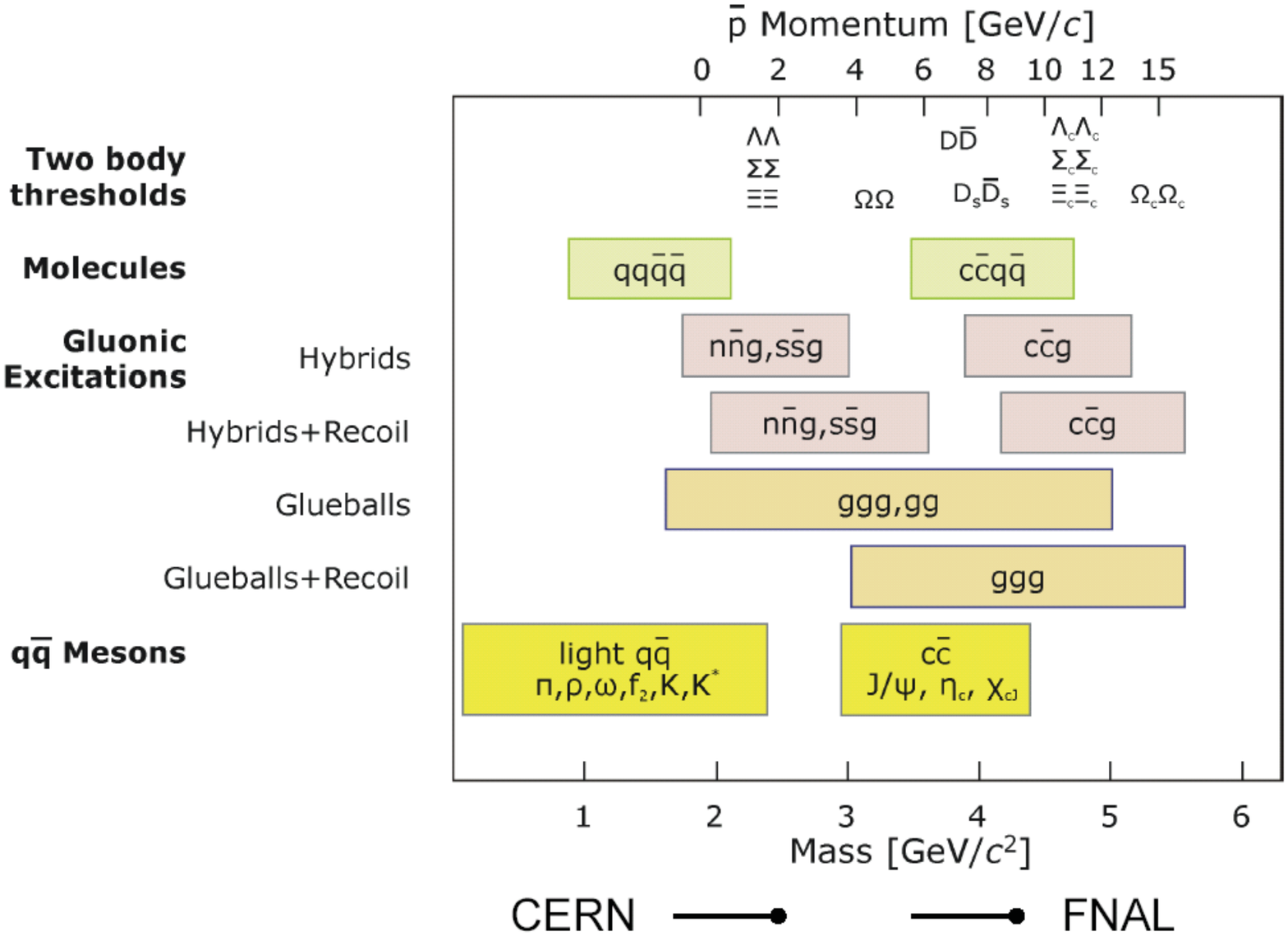}
\caption{Range of masses accessible at $\bar{P}ANDA$. Several bound systems produced in formation and production are indicated.}
\label{fig:massrange}
\end{figure}

The $\bar{P}ANDA$ (Anti\underline{P}roton \underline{An}nihilations at \underline{Da}rmstadt) experiment will be installed at the Facility for Antiproton and Ion research, FAIR, which is under construction in Darmstadt, Germany. $\bar{P}ANDA$ will use beams of antiprotons stored and accelerated in the High-Energy Storage Ring HESR. With $\bar{p}$ momenta between $1.5$ and $15$\,GeV/c, fixed-target experiments on hydrogen targets nicely cover the region of bound systems containing strange and charm quarks, as shown in Figure~\ref{fig:massrange}. The energy range exceeds the capabilities of all previous facilities greatly and will allow comprehensive studies on light-quark degrees of freedom previously begun at the LEAR antiproton storage ring at CERN, which was rather limited in beam energy, as well as the charmonium sector where experience in the exploitation of $\bar{p}p$ collisions for precision charmonium spectroscopy has been gained at FNAL.\\

%%%%%%%%%%%%%%%%%%%%%%%%%%%%%%%%%%%%%%%%%%%%%%%%%%%%%%%%%%%%%%%%%%%%%%%%%%%

\begin{figure}[htb]
\centering
\includegraphics[height=3.0in]{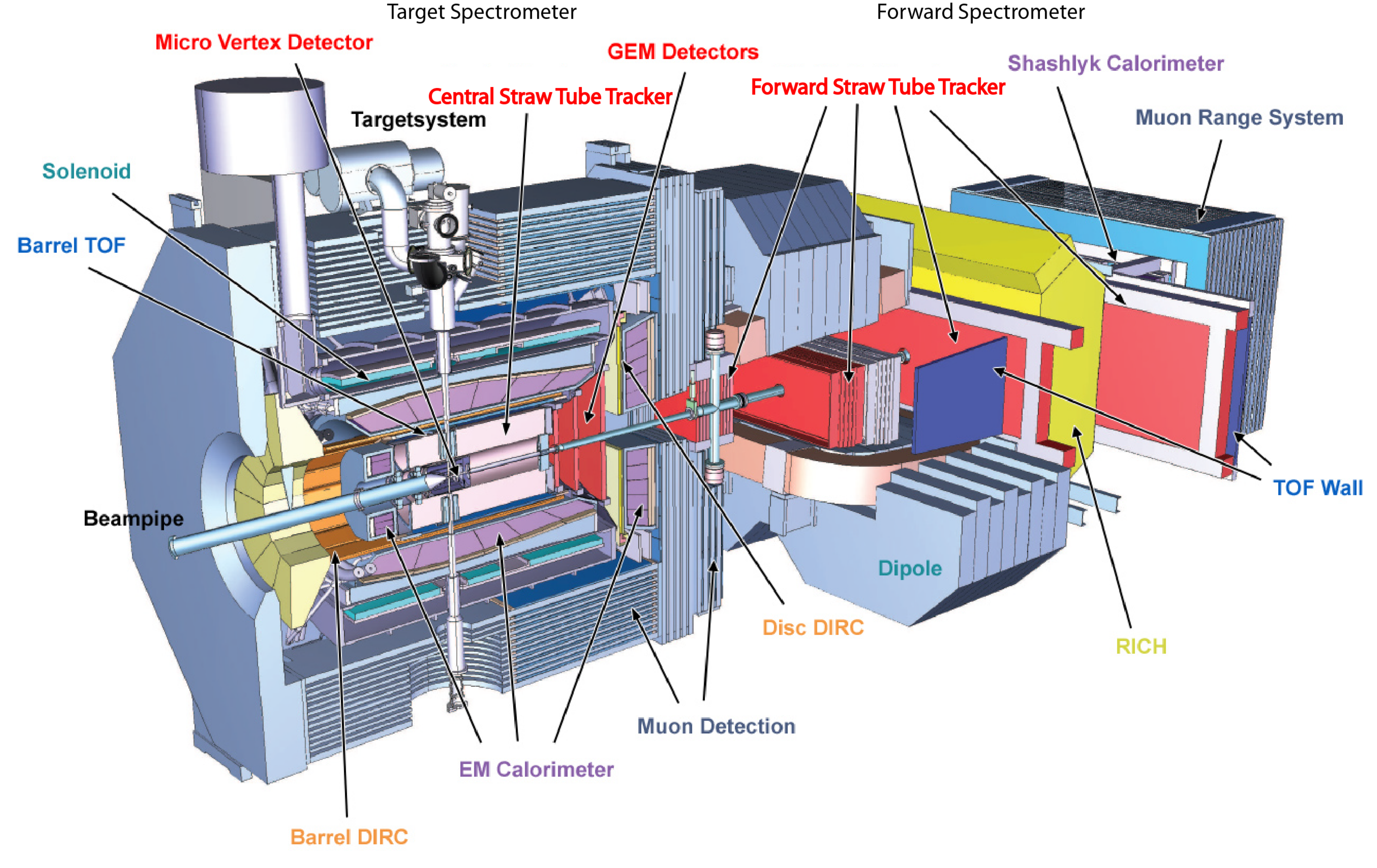}
\caption{Artist's view of the $\bar{P}ANDA$ detector with labeled sub-systems.}
\label{fig:panda}
\end{figure}

Besides the cluster-jet and pellet hydrogen target options envisaged for the $\bar{P}ANDA$ interaction region, also heavier gaseous as well as solid wire and foil targets will be used. 
$\bar{P}ANDA$ features two magnetic spectrometers to cover large transverse momenta in a solenoid surrounding the interaction region as well as forward-going particles, which will be detected in a forward dipole spectrometer. Interaction rates of up to $2\times 10^7$ per second and the large range of momenta for the outgoing particles are among the challenges that have to be addressed in the detector development. The design of the subsystems of the $\bar{P}ANDA$ spectrometers is well advanced. Prototypes of most of the detectors have been subjected to beam tests and the mechanical setup comprising all components is under study. Figure~\ref{fig:panda} shows an artist's view of the apparatus. The different components and sub-systems are indicated.

%%%%%%%%%%%%%%%%%%%%%%%%%%%%%%%%%%%%%%%%%%%%%%%%%%%%%%%%%%%%%%%%%%%%%%%%%%%
Technical design reports for a number of sub-systems are available~\cite{TDRs} after review, those for the other detector systems will follow suit.

With relatively minor modifications, the $\bar{P}ANDA$ setup can also be adapted to experiments aiming at the study of singly and doubly strange hypernuclei. In order to do this, associate production of $\Xi$ $\bar{\Xi}$ pairs in a primary target will be used. The observed decay of the $\bar{\Xi}$ will be used as a tag for implantation of the $\Xi$ in a secondary target, where it will undergo strong interaction with a nucleus and produce a hypernucleus, whose decay can then be studied.

In the following, we will restrict the discussion here to a few selected examples of advanced detector components, which feature novel properties and developments. This selection is somewhat arbitrary, but one example of each major group of detectors (particle tracking, particle ID, electromagnetic calorimetry) was chosen.

\section{Particle Tracking: Micro-Vertex Detector (MVD)}

\begin{figure}[htb]
\centering
\includegraphics[height=3.0in]{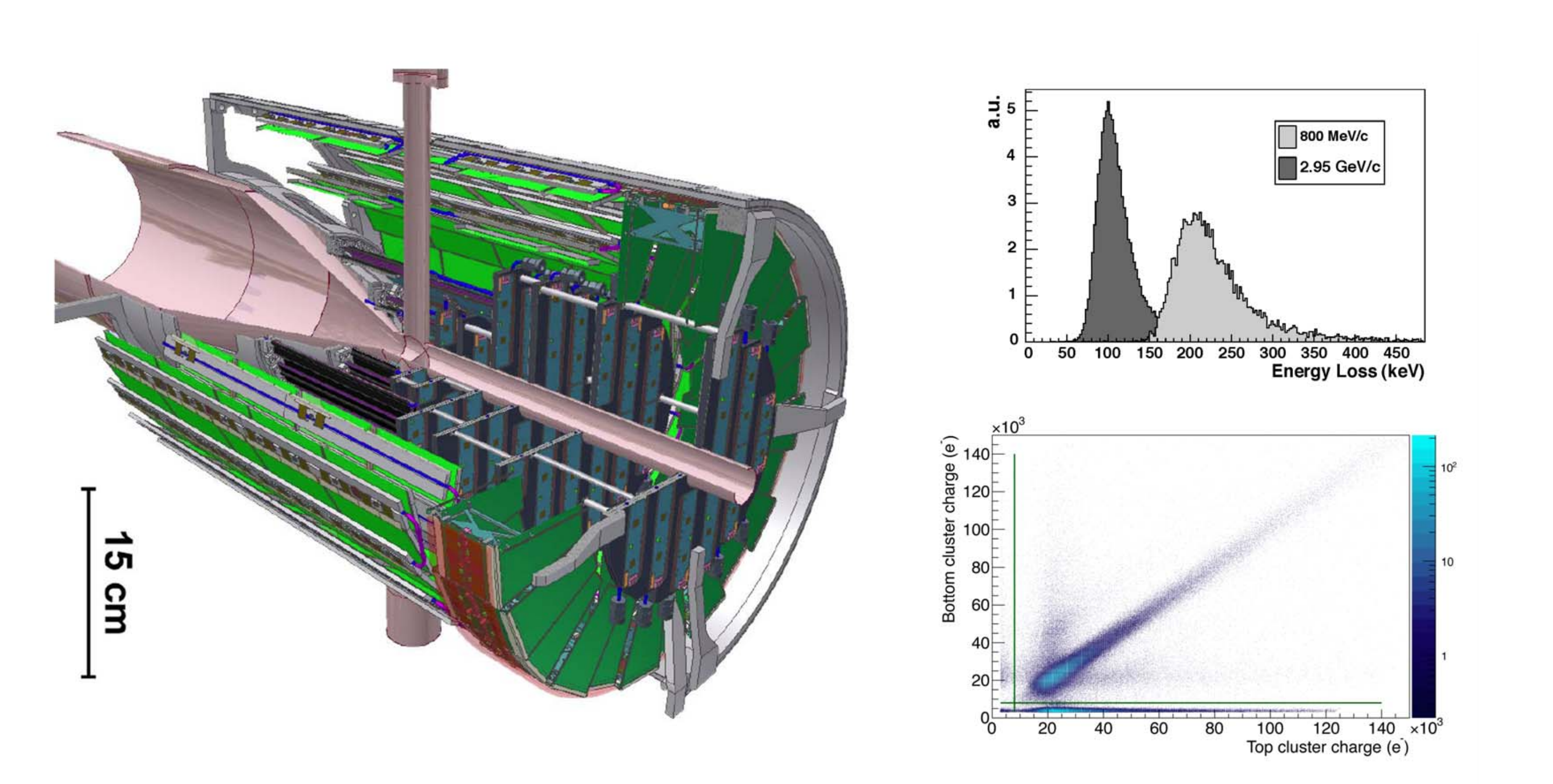}
\caption{CAD rendering of the $\bar{P}ANDA$ MVD (left) and test experiment results for a prototype double-sided strip sensor (right; top frame: charge deposit by protons of different momenta; bottom frame: correlation between the x and y strips of the sensor.}
\label{fig:MVDcombi}
\end{figure}

The innermost detector of the $\bar{P}ANDA$ setup is the Micro-Vertex Detector. It comprises a number of layers of advanced Si pixel sensors followed by Si strip detectors further outside and is designed to yield four hit layers per penetrating particle. Figure~\ref{fig:MVDcombi} shows a rendering of the very compact detector with a cut-out to reveal the layers around the beam pipe. The MVD is subdivided into forward disk and central barrel structures to account for the forward-peaked particle distribution and allow for pumping in backward direction to effectively remove residual gas from the target region. Apart from the four-hit condition, the MVD has been optimized for minimum mass budget, resolution and count rate capability at rather harsh radiation conditions. The pixel layers will be made from $100\, \mu m$ thin Si with a pixel size of $100 \times 100\,\mu m ^2$ bump-bonded to the customized readout structure of the ToPix chip that has been designed in view of the free-running data acquisition scheme of $\bar{P}ANDA$, which will do without any fast first-level trigger selections. First beam tests with detector prototypes were performed and data were successfully collected in coincidence with Si strip and scintillator detectors. The outer double-sided strip detector layers with a pitch of about $70\,\mu m$ will also be read out by a free-running front-end. Mechanical structures and the cooling scheme for the setup are under prototype testing. Some results from the beam tests performed at the COSY facility in J\"{u}lich and elsewhere can be seen to the right in Figure~\ref{fig:MVDcombi}. The top frame shows the charge distributions in a strip detector prototype measured at different beam energies, while the lower frame shows the correlation between the orthogonal strips on either side of the sensor. The correlation between both charges can be used for effective ghost hit recognition and background suppression.

%%%%%%%%%%%%%%%%%%%%%%%%%%%%%%%%%%%%%%%%%%%%%%%%%%%%%%%%%%%%%%%%%%%%%%%%%%%

\section{Particle ID: Detection of Internally Reflected Cherenkov Light, DIRCs}

Particle ID is an issue in a hadronic environment with a very large range of particle momenta spanning two orders of magnitude from a few $100\,MeV/c$ to $10\,GeV/c$. While for the particles at the lower energy boundary, energy loss and time-of-flight measurements can be used to separate pion, kaons and protons, Cherenkov detectors will be instrumental for momenta beyond $1\,GeV/c$. For the barrel section, 80 synthetic fused silica bars that form a barrel-like structure will be used to generate Cherenkov light, which is transported to the readout section with about 15 thousand MCP-PMTs by total reflection on the surfaces of each bar. With more than 20 photoelectrons per particle track and resolutions better than $10\,mrad$ for the single-photon Cherenkov angle, a very good discrimination among the various particle species can be expected and is demonstrated in simulations and test experiments.

\begin{figure}[htb]
\centering
\includegraphics[height=4.0in]{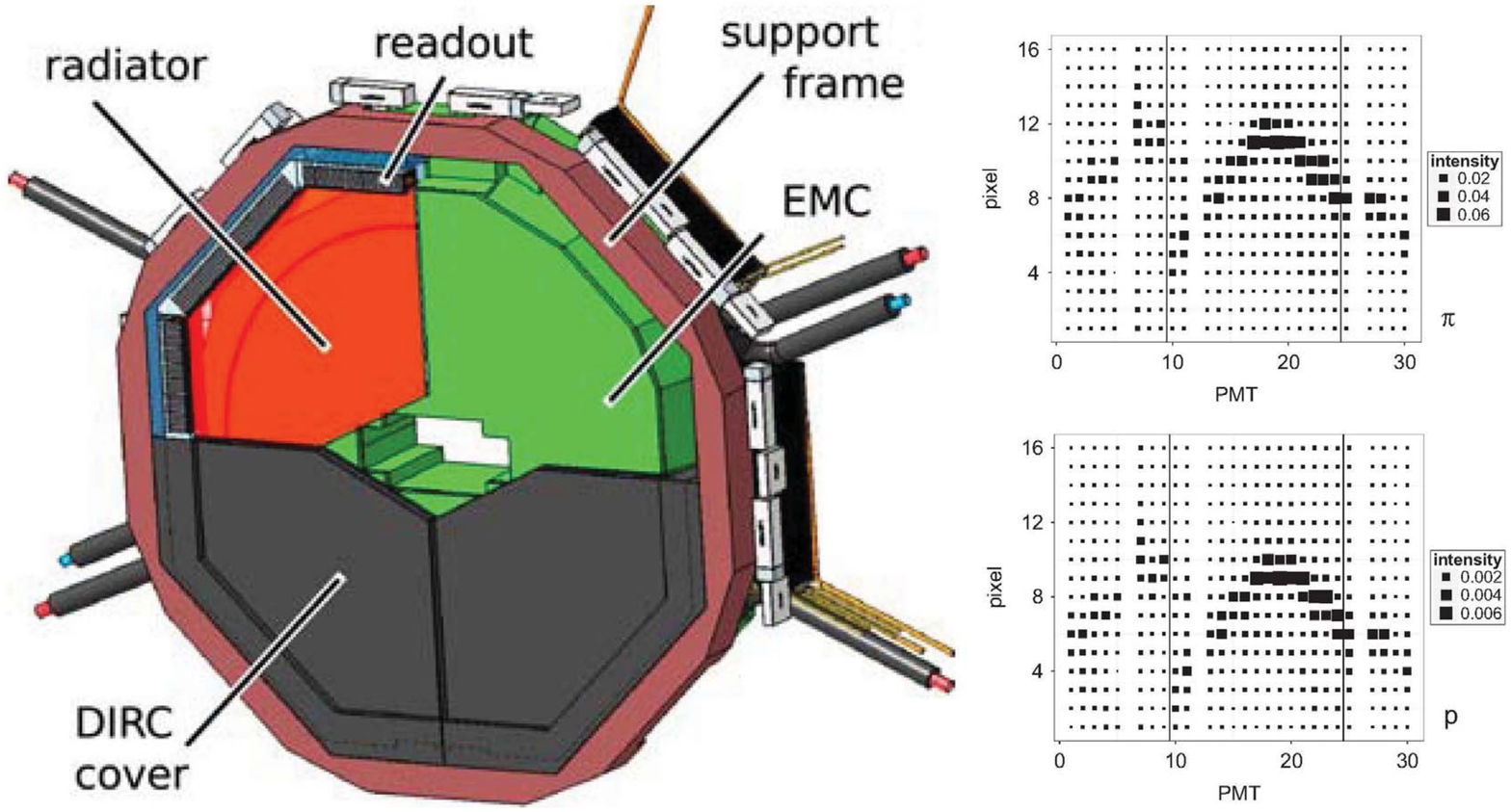}
\caption{Layout of the $\bar{P}ANDA$ disk DIRC (right) and cumulative hit patterns for pions (top) and protons (bottom) in a prototype test. The figure is adopted from~\cite{KF13}.}
\label{fig:diskDIRC}
\end{figure}

In forward direction, a disc structure with a diameter of $2\,m$ subdivided into four sectors that is equipped with dichroic mirrors along the rim, focussing light guides and readout structures using silicon photomultipliers or MCP PMTs, will yield the particle ID and velocity measurements. The overall structure of this device and the mounting in front of part of the electromagnetic calorimeter is detailed in Figure~\ref{fig:diskDIRC}. Details of the layout and performance studies can be found in~\cite{KF13}. The small frames to the right show cumulative hit patterns as measured in a test experiment with a prototype detector.The upper frame is for pions while the lower one is for protons. The proton pattern is shifted downwards with respect to that generated by pions.

%%%%%%%%%%%%%%%%%%%%%%%%%%%%%%%%%%%%%%%%%%%%%%%%%%%%%%%%%%%%%%%%%%%%%%%%%%%

\section{Electromagnetic Calorimetry}

The electromagnetic calorimeter (EMC) of $\bar{P}ANDA$ comprises about 16,000 PbWO$_4$ crystals that are grouped in backward and forward endcaps and a central barrel structure. Readout is mostly achieved using two large-area Avalanche Photodiodes, APDs, of about $1\,cm ^2$ in active surface per crystal. The EMC is operated at a temperature of $-25 ^{\circ} C$. In a series of test experiments, the performance in terms of energy and time resolution has been investigated with different probes that cover the full energy range from a few tens of MeV to 10 GeV. 

\begin{figure}[htb]
\centering
\includegraphics[height=3.0in]{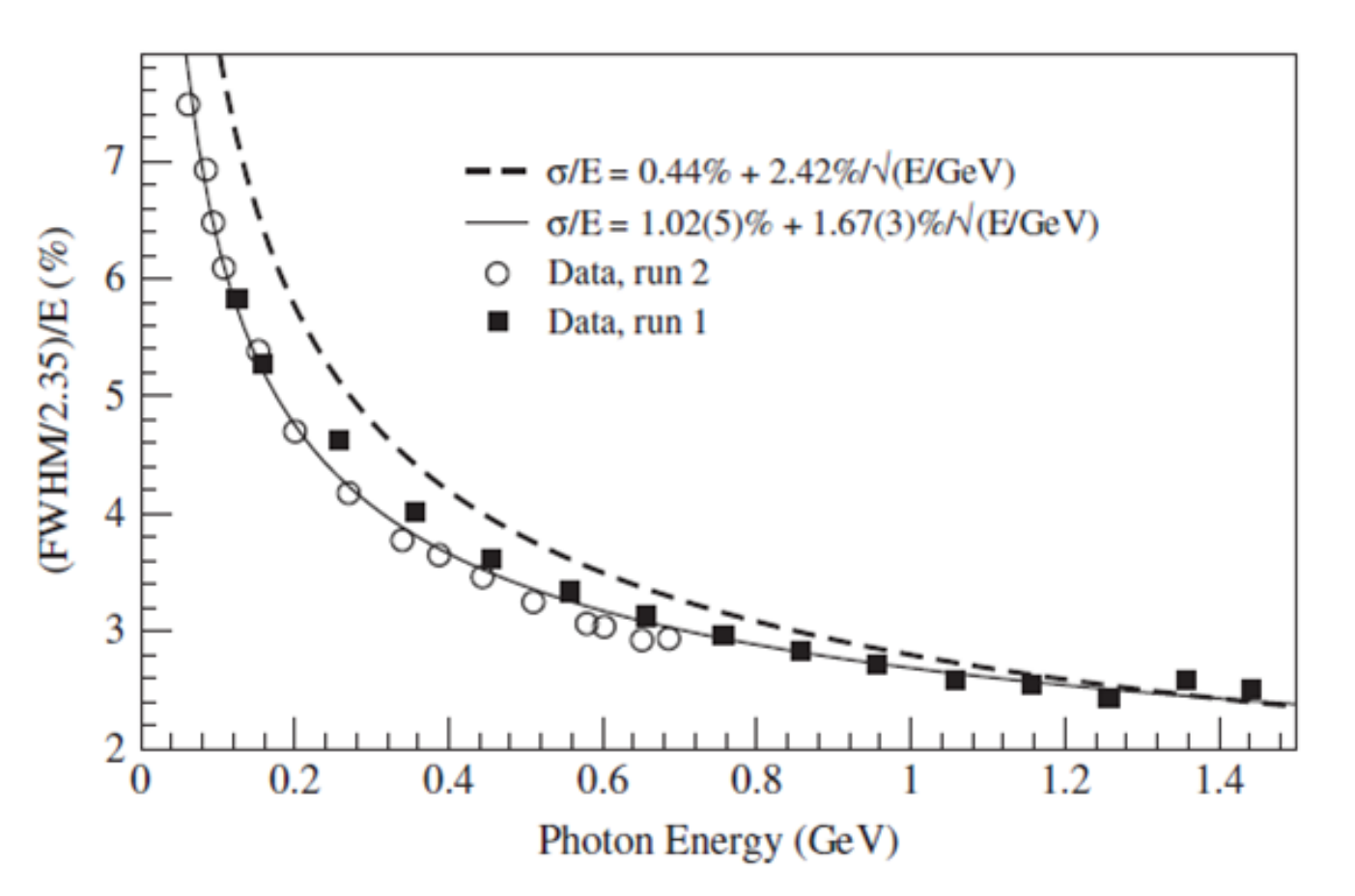}
\caption{Resolution measured with the PROTO 60 prototype of the  $\bar{P}ANDA$ EMC at the MAMI tagged photon facility.}
\label{fig:EMCresolution}
\end{figure}

As an example, Figure~\ref{fig:EMCresolution} shows a compilation of measurements using a prototype with a total of 60 crystals (PROTO 60) at the MAMI photon beam. The resolution is comparable to the design goals, while further improvements are expected from the final set of APDs as the prototype was equipped with only one smaller-than-final APD per crystal. Shown are two sets of experimental data from separate runs evaluated for a $3 \times 3$ crystal matrix as well as a fit to the data (dashed line) and the corresponding curve with standard electronics~\cite{EMC}.

%%%%%%%%%%%%%%%%%%%%%%%%%%%%%%%%%%%%%%%%%%%%%%%%%%%%%%%%%%%%%%%%%%%%%%%%%%%

%%%%%%%%%%%%%%%%%%%%%%%%%%%%%%%%%%%%%%%%%%%%%%%%%%%%%%%%%%%%%%%%%%%%%%%%%
%%
%%   use this format to include a LaTeX table  into your paper
%%
%%\begin{table}[t]
%%\begin{center}
%%\begin{tabular}{l|ccc}  
%%Patient &  Initial level($\mu$g/cc) &  w. Magnet &  
%%w. Magnet and Sound \\ \hline
%% Guglielmo B.  &   0.12     &     0.10      &     0.001  \\
%% Ferrando di N. &  0.15     &     0.11      &  $< 0.0005$ \\ \hline
%%\end{tabular}
%%\caption{Blood cyanide levels for the two patients.}
%%\label{tab:blood}
%%\end{center}
%%\end{table}
%%%%%%%%%%%%%%%%%%%%%%%%%%%%%%%%%%%%%%%%%%%%%%%%%%%%%%%%%%%%%%%%%%%%%%%%%%%

\section{Summary and outlook}

The construction of the $\bar{P}ANDA$ detector to be installed at the FAIR facility is well under way. Solutions for many of the technical challenges that are posed by the demanding environment of a high-energy antiproton beam impinging on a fixed targets of hydrogen and heavier material have been developed. While the research and development phase is still ongoing, first Technical Design Reports have been approved and many components are subjected to tests leading to final designs. According to current planing, $\bar{P}ANDA$ will be ready to see first beams in 2018. This timeline will be met by all detector components.

\Acknowledgements
The author is indebted to the BMBF for funding under project number 05P12RGFP6 and by HICforFAIR. $\bar{P}ANDA$ is supported by the national funding agencies of the participating groups.


\begin{thebibliography}{99}

%%
%%  bibliographic items can be constructed using the LaTeX format in SPIRES:
%%    see    http://www.slac.stanford.edu/spires/hep/latex.html
%%  SPIRES will also supply the CITATION line information; please include it.
%%
\bibitem{SL13}
S.\,Lange, this conference.

\bibitem{TDRs}
Technical Design Report for the PANDA Electromagnetic Calorimeter (2008), arXiv:0810.1216v1.\\
Technical Design Report for the PANDA Solenoid and Dipole Spectrometer Magnets (2009), arXiv:0907.0169.\\
Technical Design Report for the PANDA Straw Tube Tracker, Eur. Phys. J. A49 (2013), arXiv:1205.5441v2.\\
Technical Design Report for the PANDA Micro Vertex Detector (2012), arXiv:1207.6581v2.

\bibitem{KF13}
K.\,F\"{o}hl et al. on behalf of the PANDA Cherenkov group, Nucl.\,Instr.\,Meth.\,Phys.\,Res. A {\bf 732}, 346 (2013).
%%CITATION = TAADD,23,2647;%%

\bibitem{EMC} M.\,Kavatsyuk et al. on behalf of the PANDA EMC group, Nucl.\,Instr.\,Meth.\,Phys.\,Res. A {\bf 648}, 77 (2011). 

\end{thebibliography}
\end{document}